\documentclass[12pt]{article}
\usepackage[totalheight = 23cm, totalwidth = 17cm]{geometry}
\usepackage{amssymb,amsmath,amsfonts,amsbsy,graphicx}
\usepackage{color}
\usepackage{psfrag}
\usepackage{cite}

\newcommand\be{\begin{equation}}
\newcommand\ee{\end{equation}}
\newcommand\ba{\begin{eqnarray}}
\newcommand\ea{\end{eqnarray}}\newcommand\eq{\begin{equation}}           
\newcommand\en{\end{equation}}

 \newcount\colveccount
\newcommand*\colvec[1]{
        \global\colveccount#1
        \begin{pmatrix}
        \colvecnext
}
\def\colvecnext#1{
        #1
        \global\advance\colveccount-1
        \ifnum\colveccount>0
                \\
                \expandafter\colvecnext
        \else
                \end{pmatrix}
        \fi
}
\input epsf

\usepackage{ulem}
\usepackage{amssymb}
\usepackage{amsmath}
\usepackage{bm}
\usepackage{graphicx}
\usepackage{color}
\usepackage{subfig}
\usepackage{caption}

\def\gsim{\;\rlap{\lower 2.5pt
 \hbox{$\sim$}}\raise 1.5pt\hbox{$>$}\;}
\def\lsim{\;\rlap{\lower 2.5pt
 \hbox{$\sim$}}\raise 1.5pt\hbox{$<$}\;}

\usepackage{graphicx}
\usepackage{url}
\usepackage{amssymb,amsmath}
\usepackage{amsbsy}

\usepackage{graphicx}
\usepackage{url}
\usepackage{amssymb,amsmath}
\usepackage{amsbsy}

\begin{document}
\title{
  The radio telescope search for the resonant conversion of cold dark matter axions from the magnetized astrophysical sources
}
 \author{Fa Peng Huang$^1$, Kenji Kadota$^1$, Toyokazu Sekiguchi$^2$ and Hiroyuki Tashiro$^3$ \\
{ \small $^1$ \it  Center for Theoretical Physics of the Universe, Institute for Basic Science (IBS), Daejeon 34051, Korea} \\
{ \small $^2$ \it Research Center for the Early Universe (RESCEU),
  Graduate School of Science,}\\
{\small The University of Tokyo, Tokyo 113-0033, Japan }\\
{ \small $^3$  \it Department of Physics, Graduate School of Science, Nagoya University, Aichi 464-8602, Japan}
}
\date{\vspace{-5ex}}
\maketitle   

\begin{abstract}
  We study the conditions for the adiabatic resonant conversion of the cold dark matter (CDM) axions into photons in existence of the astrophysically sourced strong magnetic fields such as those in the neutron star magnetosphere. We demonstrate the possibility that the forthcoming radio telescopes such as the SKA (Square Kilometre Array) can probe those photon signals from the CDM axions. 
\end{abstract}

\setcounter{footnote}{0} 
\setcounter{page}{1}\setcounter{section}{0} \setcounter{subsection}{0}
\setcounter{subsubsection}{0}

\section{Introduction}

Since the proposal of Peccei-Quinn (PQ) mechanism for an elegant solution of the QCD strong CP problem, there have been many attempts to search for the axion which naturally arises as a pseudo-scalar particle of the PQ symmetry \cite{pe1977,we1977,wil1977,kim1979,shi1979,dine1981,zhi1980}.
Besides the QCD axions, more generally, the axion-like particles (ALPs) also have been widely discussed which can commonly arise in the string theory \cite{sv2006}.
The possibility for these axions/ALPs to be the CDM candidates also gives a tantalizing motivation to search for them \cite{sik2006,arias2012,ring2012,gra2015}.
It is intriguing that the axion CDM mass range $\mu eV \sim m eV$ (corresponding to the frequency $ 0.1 GHz \sim 100 GHz$) which is motivated from the QCD axion as a CDM candidate turns out to overlap with the frequency range which the radio telescope can probe \cite{kap1985,tur1989,bor2016}.

We seek the radio telescope probe of CDM axions through their adiabatic resonant conversion into photons in the astrophysically sourced strong magnetic fields such as those in the vicinity of the neutrons stars/magnetars. This is in stark contrast to the relativistic axion with the x-ray energy for which it has been claimed that the adiabatic resonant conversion cannot be realized in the strongly magnetized plasma such as in the neutron star magnetosphere due to the significant vacuum polarization contribution to the photon dispersion relation \cite{raf1987,yoshi1988}.

The axion and photon can convert to each other in existence of the magnetic fields through the Primakoff process and many attempts have been made to seek the axions using a powerful magnet in the laboratory to result in the tight bounds on the axion mass and its coupling to photons \cite{sik1983,wu1989,de1987,hag1990,as2009,ana2017,pat2016}. Many studies also have been done for the axion search using the astrophysically sourced magnetic fields such as the intergalactic magnetic fields and stellar magnetic fields \cite{raf1987,yoshi1988,yana1987,hoch2007,miriraf2006,sch2015,miri2009b,higaki2013}.
The use of actual astrophysical data from the gamma ray, X-ray, optical and radio telescopes also helped in reducing the viable axion parameter space, but many of those analysis assumed the relativistic axion converting into photon or the CDM axion decaying into two photons \cite{fermi2016,be2016,grin2006,bl2000}.
The potential radio telescope probe of the non-relativistic axion converted into the photon in presence of the astrophysical magnetic fields have been recently studied assuming the non-resonant conversion and little study has been done for the resonant conversion for the radio surveys \cite{kel2017,sigl2017,bai2017,psh2007}. Our study on the adiabatic resonant conversion of CDM axion would complement those previous studies for the further radio telescope exploration of axion search. \S 2 outlines the setup of our study and \S 3 examines the conditions for the adiabatic resonant conversion of axions into photons. \S 4 discusses the detectability of the photon flux by a radio telescope as a result of such an efficient axion-photon conversion.

\section{The axion-photon wave propagation in the magnetic fields}
The Lagrangian for the axion-photon system in existence of the magnetic fields relevant for the magnetized astrophysical sources such as the neutron stars is
\ba
L=-\frac{1}{4}F_{\mu\nu}F^{\mu\nu}
+\frac{1}{2}(\partial_\mu a \partial^\mu a -m_a^2 a^2)+L_{int}+L_{QED}~,
\ea
where $a$ is the axion with the mass $m_a$ and $F_{\mu \nu}$ is the electromagnetic field tensor. The pseudo-scalar axion can convert to the spin-1 photon in existence of the external magnetic field perpendicular to the photon propagation, and the interaction term in the Lagrangian for the electromagnetic field and the axion is
\ba
L_{int}=\frac{1}{4} g \tilde{F}^{\mu \nu }F_{\mu \nu} a=-g {\bf E} \cdot {\bf B} a
\ea
where $g$ represents the axion-photon coupling with the dimension [mass]$^{-1}$ and $E$ is the electric field associated with the photon and $B$ is the transverse component (with respect to the photon propagation) of the magnetic field \footnote{The photon here has a liner polarization parallel to the external magnetic field. The other photon polarization state and the photon mass term due to the Cotton-Mouton effect (which can cause the birefringence) are of little importance in our discussions and thus will be ignored \cite{adler1971,raf1987}.}. The axion in our discussions, for the sake of brevity, refers to the axion and more generally to the ALP as well defined by this Lagrangian characterized by its mass and coupling to the photon (we accordingly treat $m_a, g$ as independent parameters).

$L_{QED}$ represents the quantum correction to the Maxwell equation (due to the QED vacuum polarization) and it can be given by the Euler-Heisenberg action whose leading order term is \cite{adler1971,raf1987}
\ba
L_{QED}=\frac{\alpha^2}{90 m_e^4} \frac{7}{4}(F_{\mu\nu}\tilde F^{\mu\nu})^2~,
\ea
where $\alpha=e^2/4\pi$ is the fine-structure constant.
The photon obtains the effective mass in the magnetized plasma. The contribution of the photon mass $m_{\gamma}^2=Q_{pl}-Q_{QED}$ come from the vacuum polarization
\ba
Q_{QED}=\frac{7 \alpha}{45 \pi} \omega^2 \frac{B^2}{B^2_{crit}}
\ea
with $B_{crit}=m_e^2/e=4.4\times 10^{13}G$ and the plasma mass characterized by the plasma frequency $\omega_{pl}$ 
\ba
Q_{plasma}=\omega_{plasma}^2=4 \pi \alpha \frac{n_e}{m_e}
\ea
with the charged plasma density $n_e$. 
It has been pointed out that the QED vacuum polarization effect spoils the realization of the adiabatic resonant conversion between the relativistic axion (with the observable x-ray energy range) and the photon in the vicinity of a neutron star with strong magnetic fields \cite{raf1987,yoshi1988}. We note here that the vacuum polarization effect is not important compared with the plasma effect for our axion CDM scenario. As a simple estimation, adopting the Goldreich-Julian charge density \cite{gol1969} for the plasma density
\ba
n_e^{GJ}=7\times 10^{-2}\frac{1s}{P}\frac{B(r)}{1G}\frac{1}{cm^3}
\ea
where $P$ is the neutron star spin period,
\ba
\frac{Q_{pl}}{Q_{QED}}
    \sim 5\times 10^{8}
     \left( \frac{\mu eV}{\omega} \right)^2
          \frac{10^{12}G}{B}
     \frac{1 sec}{P}~.
\ea
We can hence safely ignore $Q_{QED}$ with respect to $Q_{pl}$ for the parameter range of our interest, because of a small photon frequency $\omega$ relevant for the frequency range sensitive to the radio telescopes in our CDM axion scenario ($\omega \sim m_a$).

The equation for the axion-photon plane wave with a frequency $\omega$ reads
\ba
\left[
\omega^2
+\partial^2_z
+
\left( \begin{array}{cc}
- m_{\gamma}^2 & gB\omega  \\
gB\omega & - m_a^2  \\
 \end{array} \right)
\right]
\left(
\begin{array}{c}
  \gamma\\
  a
  \end{array}
\right)
=0~,
\ea
where we assumed for simplicity the time-independent magnetic field $B(r)$ \cite{raf1987}. The mass matrix here can be diagonalized by the rotation unitary matrix
\ba
U=
\left( \begin{array}{cc}
\cos \tilde{\theta} & \sin \tilde{\theta}  \\
-\sin \tilde{\theta} & \cos \tilde{\theta}  \\
 \end{array} \right)
\ea
with
\ba
\label{mixang}
\cos 2 \tilde{\theta}=
\frac{m_a^2-m_{\gamma}^2}{\sqrt{4 g^2 B^2 \omega^2+(m_{\gamma}^2-m_a^2)^2}}
\nonumber
\ea
\ba
\sin 2 \tilde{\theta}=
\frac{2 gB\omega}{\sqrt{4 g^2 B^2 \omega^2+(m_{\gamma}^2-m_a^2)^2}}
\label{mixingangle}~,
\ea
where the tilde represents the mixing angle in the medium to be distinguished from that in the vacuum. The maximum mixing can occur when $m_{\gamma}^2(r)\approx m_{a}$. 
The mass eigenvalues are
\ba
m_{1,2}^2=
\frac{
  (m_{\gamma}^2+m_a^2)
  \pm
  \sqrt{
(m_{\gamma}^2-m_a^2)^2+4g^2 B^2 \omega^2
  }
}
     {2}
\ea
with the corresponding momentum for the mass eigenstates
\ba
k_{1,2}^2=\omega^2-m_{1,2}^2~.
\ea


If the magnetic field is homogeneous, the conversion probability for the axion into photon becomes 
\ba
p_{a\rightarrow \gamma}=\sin ^2 2 \tilde{\theta}(z) \sin^2 [z(k_1-k_2)/2]
\ea
for the wave dominated by the axion component at $z=0$.
This is analogous to the neutrino oscillations and we can interpret the axion-photon conversion in an analogous manner. Even though the magnetic field is inhomogeneous in the neutron star magnetosphere, the conversion in such a non-uniform magnetic field can be studied analogously to the MSW effect for the neutrino oscillations in the spatially varying matter background \cite{wo1977,mik1986}. The wave initially dominated by the axion component can maximally mix with the photon in the resonance region in existence with the strong magnetic fields and it gets adiabatically transformed into the photon state, resulting in the photon dominated wave outside the magnetosphere.

We now more quantitatively discuss the conditions for the adiabatic resonant conversion of CDM axion into photons.
\section{The adiabatic resonant conversion of axions into photons}
\label{secadi}
The resonance can occur when the maximum mixing angle is realized for $m_{\gamma}^2(r) \approx m_a^2$.
The photon mass or the plasma mass depends on the plasma density. The realistic modeling of the magnetosphere of a neutron star is beyond the scope of this paper, and we simply assume a simple dipole magnetic field with a magnitude at the neutron star surface $B_0$ and the charged plasma density obeying the Goldreich-Julian density 
\ba
\label{dipole}
B(r)=B_0
\left(
\frac{r}{r_0}
\right)^{-3}
\ea
and
\ba
\label{edensity}
m^2_{\gamma}(r)=
4 \pi \alpha \frac{n_e(r)}{m_e},
n_e(r)= n_e^{GJ}(r)=7\times 10^{-2}
\frac{1s}{P}\frac{B(r)}{1G}\frac{1}{cm^3}
\ea
where $r_0$ is the neutron star radius.
The resonance radius is defined at the level crossing point $m^2_{\gamma}(r_{res})=m_a^2$ given by
  \ba
  \label{resr}
   \left(
   \frac{r_{res}}{r_0}\right)^{-3}
   \approx
 10^{-3}
   \left(
   \frac{m_a}{\mu eV}
   \right)^2
   \left(
   \frac{10^{14}G}{B_0}
   \right)
      \left(
   \frac{P}{10 sec}
   \right)  ~.
   \ea
At the resonance $|m_{\gamma}^2-m_a^2|\ll gB\omega$ and $m_{1,2}^2 \approx m_a^2\pm gB\omega$.
From the mixing angle given in Eq. (\ref{mixang})
 \ba
\sin 2 \tilde{\theta}=
\frac{\left(2 gB\omega/m_{\gamma}^{2}\right)}
     {\sqrt{\left(4 g^2 B^2 \omega^2/m_{\gamma}^4\right)+\left(1-\left(m_a/m_{\gamma}\right)^2 \right)^2}}
\equiv
\frac{c_1}{\sqrt{c_1^2+(1-f(r))^2}}
\ea
where $c_1$ is a constant independent of the radius, we can see that the resonant occurs when $f(r)\equiv \left( m_a/m_{\gamma} \right)^2=1$ with the resonance width $\Gamma=2 c_1\equiv 4 g B \omega/m_{\gamma}^2$.

We first examine the adiabatic condition for the sufficient conversion of axions.
The adiabatic resonant conversion requires the region in which the resonance is approximately valid inside the resonance width
\ba
\delta r \sim \delta f \left| df/dr \right|_{res}^{-1} \sim 2 c_1 \left| df/dr \right|_{res}^{-1}
\ea
is sufficiently bigger than the oscillation length scale at the resonance
\ba
l_{osc}=\frac{2\pi}{|k_1-k_2|_{res}}~.
\ea
$\delta r > l_{osc}$ hence requires
\ba
 \left|
d \ln f/dr
\right|^{-1}_{res}
>
650 [m]
\left(
\frac{m_a}{\mu eV}
\right)^3
\left(
\frac{v_{res}}{10^{-1}}
\right)
\left(
\frac{1/10^{10}GeV}
     {g}
\right)^2
\left(
\frac{10^{12}G}{B(r_{res})}
\right)^2
\left(
\frac{\mu eV}{\omega}
\right)^2~.
\ea
The velocity at the resonance $v_{res}$ can be affected by the gravitational acceleration near the neutron star and can be much bigger than the characteristic CDM velocity in our solar neighborhood $v\sim 10^{-3}$ (e.g. the escape velocity can be of order $v\sim {\cal O}(0.1)$ inside the magnetosphere of a neutron star).
This adiabaticity condition means the scale relevant for the plasma density variation should be bigger than the scale indicated on the right-hand side. The typical scale for the magnetosphere (or the Alfven radius) is or order $100 r_0\sim {\cal O}(10^6) m$ and we can infer that this variation length scale required for the adiabaticity can well be within the neutron star magnetosphere. 
This condition is equivalent to $|{d \tilde{\theta}}/{dr}|_{res}<  l_{osc}^{-1}$ as readily checked by using Eq.(\ref{mixingangle}) and the resonance condition $m_{\gamma}^2=m_a^2$. The adiabatic condition hence assures us that the mixing angle variation is slow enough assuming that the density variation is sufficiently smooth so that the higher order terms do not become significant. 

For the axion-photon wave propagation in the magnetosphere, due to the existence of the plasma medium, we also demand the coherence of the wave propagation for the resonant conversion. This gives additional constraints which do not show up for the analysis of the conventional neutrino oscillations. The incoherent scatterings between the converted photon and plasma medium, such as the Thomson scatterings, can lose the coherence of the wave propagation \cite{raf1987,yoshi1988}. We demand the photon mean free path exceeds the oscillation length to prevent the photon component of the beam from incoherently scattering with the plasma. The Thomson scattering
\ba
\sigma=8 \pi \alpha^2 / 3 m_e^2 ~,
\ea
and the mean free path is
\ba
\frac{1}{\sigma n_e} \sim
\frac{10^7 km}{n_e/(10^{12}/cm^3)} 
\ea
which exceeds all the relevant length scales of our discussions ($n_e\sim 10^{12}/cm^3$ corresponds to the gas density at the neutron star surface) and hence does not affect our discussions \footnote{The strong magnetic fields can possibly affect the Thomson scattering cross section, which however does not lead to the violation of this coherence condition for the parameter range of our interest \cite{can1971,her1979}.}.
We also require the photon effective refractive index to be real
\ba
n_{1,2}^2=1-\frac{m_{1,2}^2}{\omega^2}=\frac{k^2_{1,2}}{\omega^2}>0
\ea
to avoid the loss of coherence in the axion-photon oscillation and the the attenuation of the wave propagation.

\section{The photon flux search by the radio telescope}
To estimate the photon flux, let us start by considering the axion particle trajectory with the initial velocity $v_{0}$ far away from the neutron star in the Schwarzschild metric. The impact parameter $b$, whose closest approach to the neutron star is $R$, is given by
\ba
b(R)=R \frac{v_{esc}(R)}{v_0}\left( 1-2 GM/R\right)^{-1/2}~,
\ea
where $M$ is the neutron star mass and $v_{esc}=(2GM/R)^{1/2}$.
  Recalling our discussion on the adiabatic resonance in \S\ref{secadi} (the efficient conversion can occur for $m_{\gamma}^2 (r_{res}) \approx m_a^2$ with the resonance width $\Delta m^2_{\gamma} \approx 4 g B \omega$), we can estimate that the axion mass going through the efficient axion-photon conversion region is of order
  \cite{press1985,gold1989,kou2007,cape2013} 
 \ba
  \frac{d m_a}{d t}\sim \pi (b^2(r_{+})-b^2(r_{-}) ) \rho_a v_0
\sim  \frac{8 \pi}{3} r_{res}GM v_0^{-1} \rho_a gB\omega m_a^{-2}
  \ea
 where $\rho_a$ is the axion CDM density and we used $gB\omega < m_{a}^2$ for the parameter range of our interest. $r_{\pm}$ is defined by $m^2_{\gamma}(r_{\pm})=m_a^2 \mp \Delta m^2_{\gamma}/2$, and we, for a conservative estimation, do not count the axions going through $r<r_{-}$ to avoid the wave attenuation. 
The photon energy from the axion-photon conversion is
\ba
 \frac{d E}{d t}=p_{a\rightarrow \gamma} \frac{dm_a}{dt}
\ea
where the conversion probability $p$ can be close to unity for the adiabatic resonant conversion, and the photon flux density can be estimated to be of order
\ba
\label{pflux}
S_{\gamma}&=&\frac{ dE/dt }
{4 \pi d^2 \Delta \nu}
\\
\nonumber
&\sim&    4.2 \mu Jy 
\frac{
\left(\frac{r_{res}}{100km}\right)
\left(\frac{M}{M_{sun}}\right)
\left(\frac{\rho_a}{0.3 GeV/cm^3}\right)
\left(\frac{10^{-3}}{v_0}\right)
  \left(
     \frac{g}{1/10^{10}GeV}
     \right)
       \left(
     \frac{B(r_{res})}{10^{12}G}
     \right)
       \left(
     \frac{\omega}{\mu eV}
     \right)
       \left(
     \frac{\mu eV}{m_a}
     \right)^2
}
     {
       \left( \frac{d}{1kpc} \right)^2
       \left(\frac{m_a/2\pi}{\mu eV/2\pi}\right) \left(\frac{v_{dis}}{10^{-3}}\right)
     }
   \ea
where $d$ represents the distance from the neutron star to us. The photon flux peaks around the frequency $\nu_{peak}\sim m_a/2\pi$ and $\Delta \nu \sim \nu_{peak} v_{dis}$ represents the spectral line broadening around this peak frequency due to the DM velocity dispersion $v_{dis}$.


We are interested in the detectability of this photon flux as a result of axion-photon resonant conversion by a radio telescope.
For this purpose, one can consider the antenna temperature induced by the total flux density $S$ 
\ba
\label{tann}
  T=\frac{1}{2}A_{eff}S
  \ea
where $A_{eff}$ represents the effective collecting area of the telescope \cite{con2016}. The minimum detectable brightness temperature (sensitivity) can be given by the root mean square noise temperature of the system (which consists of the added sky/instrumental noises of the system)
\ba
\label{tmin}
   T_{min} \approx \frac{T_{sys}}{ \sqrt{\Delta B t_{obs}}}~,
  \ea
where $\Delta B$ is the bandwidth and $t_{obs}$ is the integrated observation time. 
  We can hence estimate, from Eqs. (\ref{tann},\ref{tmin}), that the smallest detectable flux density is of order
  \ba
    \label{smin}
  S_{min} \approx  0.29 \mu Jy  
\left(
\frac{1 GHz}{\Delta B}
\right)^{1/2}
\left(
\frac{24 hrs}{t_{obs}}
\right)^{1/2}
\left(
\frac{10^3 m^2/K}{A_{eff}/T_{sys}}
\right)
\ea
to be compared with the photon flux from the axion conversion given by Eq. (\ref{pflux}). $A_{eff}/T_{sys}$ differs for different experiment specifications. For instance, the SKA-mid in the Phase 1 (SKA1) will be able to provide $A_{eff}/T_{sys}\sim 2.7 \times 10^3 $ assuming $A_{eff} \sim (180m)^2$ and $T_{sys}\sim 12K$, and it would increase by more than an order of magnitude assuming $A_{eff}\sim (1km)^2$ in the Phase 2 (SKA2) \cite{ska}.

There still exists a wide range of axion parameter space of $m_a,g$ which still has not been explored, and the radio telescope can complement the other experiments to fill in the gap of those unexplored parameter spaces.
For instance the FAST (Five hundred-meter Aperture Spherical radio Telescope) covers $ 70MHz \sim 3GHz$, the SKA (Square Kilometre Array) for $50MHz\sim 14 GHz$ and the GBT(Green Bank Telescope) for $0.3 \sim 100 GHz$, so that the radio telescopes can probe the axion mass range of $0.2 \sim 400 \mu eV$ \cite{fast,ska,gbt}. The current axion search experimental upper bounds on the axion-photon coupling corresponding to this radio telescope frequency range is $g<6.6 \times 10^{-11} GeV^{-1}$, which comes from the helioscope experiment 
and also from the energy loss rate enhancement of the horizontal branch stars of global clusters through the Primakoff effect \cite{ana2017,aya2014}.
The haloscope (microwave cavity) experiments give even tighter bounds for some limited axion mass ranges. For instance, $g\lesssim 10^{-15} GeV^{-1}$ for the axion mass of $2 \sim 3.5 \mu eV $ and $g\lesssim 10^{-13} GeV^{-1}$ for the axion mass of $ 4.5 \sim 10 \mu eV$ \cite{wu1989,hag1990,as2009,pat2016,ry2017}.
\\
The exclusive parameter search for our study on the radio telescope probe for axions is beyond the scope of this paper partly because of the astrophysical uncertainties in the magnetosphere modeling and a wide range of the possible parameters for the neutron stars (e.g. the spin period can vary in a wide range (${\cal O}(10^{-3}\sim 10^3)$ seconds) and the magnetic field can reach up to $10^{15}$G) \cite{mil2014,ho2013,yan2017,jo2017,tau2014,klus2013,shaku2011,turolla2015,magcata}). Dark matter properties such as the dark matter velocity dispersion in the neighborhood of a neutron star remain to be clarified too. The dark matter density in the vicinity of a neutron star does not have to be same as that in the solar neighborhood $\rho_a\sim 0.3 GeV/cm^3$. For instance, in the region where the neutron star distribution peaks in our galaxy ($\sim$ a few kpc from the galactic center), the density can well be enhanced by more than an order of magnitude (e.g. $\rho_a \sim {\cal O}(10\sim 100) \times 0.3 GeV/cm^3$) and could be even bigger $\rho_a \sim {\cal O}(10^4) \times 0.3 GeV/cm^3 $ around the neutron star found near the galactic center due to a dark matter spike \cite{lori2006,cerm2017,re2014,meri2002,whar2011,den2009,eat2013}.

For a trial parameter set, let us adopt the DM velocity and the dispersion velocity of order $v_0\sim v_{dis}\sim 10^{-3}$ and a factor 10 enhancement of the local DM density compared with the value near the earth $\rho_a\sim 3 GeV/cm^3$. Let us also assume a neutron star of order a kpc away from us and take the DM velocity in the resonance region of order the escape velocity at the resonance radius. Then, for our toy magnetosphere model with a simple dipole magnetic field profile (Eqs. (\ref{dipole},\ref{edensity})), a parameter set ($B_0= 10^{15}G,m_a=50 \mu eV,P=10s,g=5 \times 10^{-11}GeV^{-1},r_0=10km,M=1.5 M_{sun}$) satisfies the conditions for the adiabatic resonance conditions with $S_{\gamma}\sim 0.51 \mu Jy$. This can exceed the estimated minimum required flux $S_{min}\sim 0.48 \mu Jy$ for the SKA1 and $S_{min}\sim 0.016 Jy$ for the SKA2 with 100 hour observation time, where we assumed the optimized band width matching the signal width $\Delta B=\Delta \nu$. This simple parameter set example would work as an existence proof for the motivation to seek a potential radio telescope probe of the adiabatic resonant conversion of axions. Even though the further parameter search with more detailed astrophysical model setups is left for the future work, we remark a few comments regarding the astrophysical uncertainties involved in our estimation before concluding our work.
\\
The measurements of the neutron star radiation in the different wavebands have been fitted well by assuming the magnetic field profile more complicated than a simple vacuum dipole profile (e.g. twisted magnetosphere) and the plasma charge density larger (e.g. a few orders of magnitude larger) than the classical Goldreich-Julian value \cite{thom2001,lyu2005,bel2006,rea2008,turolla2015,ak2016,chen2016,kas2017}. Such an enhancement in the plasma density would increase the resonance radius and can affect the adiabaticity condition and the photon flux estimation. In addition to the DM velocities and velocity dispersions which can affect our photon flux estimation, the galactic drift velocities of neutron stars are also uncertain parameters whose velocity distribution does not follow a simple Gaussian-like distribution and spans in a wide range (typically ${\cal O}(100\sim 500) km/s$ with a significant fraction ($\sim {\cal O}(10) $\%) of the neutron star population having the velocity exceeding $1000 km/s$) \cite{ar2001}. Such variations in the relative velocity between a neutron star and the axions can affect the photon flux estimation too. A more detailed study taking account of such astrophysical uncertainties and the numerical analysis for our adiabatic resonance conversion scenario along with the extension of our scenario to the one including the non-adiabatic axion-photon conversion are left for our future work.


Let us here also briefly comment on the comparison of our scenario with the other relevant works. The conditions of the complete conversion of axions into photons were first studied in \cite{yoshi1988} which however considered the relativistic axions with the X-ray energy range, in contrast to the radio range in our scenario, and hence could not realize the sufficient adiabaticity due to the QED effect. Our CDM axion scenario is also in contrast to the resonant conversion scenarios where the only partial axion-photon conversion, hence with a smaller conversion probability compared with the complete conversion scenarios, occurs due to the insufficient adiabaticity and/or the coherence. Such a partial conversion scenario would relax the bounds on the model parameters and would be applicable for a larger sample of the neutron stars, but one needs to require some way to compensate a smaller conversion probability to detect the axion signals by the radio telescopes such as a large dark matter density near the neutron star (e.g. the dark matter density enhancement factor of ${\cal O}(10^{10})$ with respect to the local density in the solar neighborhood \cite{hook} \footnote{Ref.\cite{hook} assumes the signal bandwidth broadening of order the DM velocity squared, in contrast to ours and the other literature which take account of a more dominant line width linear in the velocity (e.g. from the Doppler broadening effect). This can result in the overestimation of the signal flux by three to four orders of magnitude and we quoted here the required dark matter density enhancement factor bigger than their adopted value taking account of this correction.}). How the model parameters are affected by the astrophysical uncertainties would differ depending on the scenarios too. For instance, for the partial conversion scenarios, a smaller $g$ would become viable if we allow a bigger DM density enhancement because a bigger dark matter density (to which the photon flux is proportional) could compensate for a smaller conversion probability. For the complete conversion scenarios, on the other hand, the tight bounds also come from, in addition to demanding the large enough signal flux, the adiabaticity and coherence conditions which have the non-trivial dependence on the plasma density rather than the dark matter density for a given dark matter mass. A smaller value of $g$ hence does not necessarily become viable just by changing the dark matter density profile to justify the large dark matter density enhancement around the neutron star. 

We demonstrated in this paper the possibility for the radio telescope detection of axions from the MSW-like resonant transition where the complete axion-photon conversion can occur, which requires the adiabaticity and the coherence resulting in the stringent constraints on the model parameters. While we studied the conditions for the adiabatic resonant conversion, the more precise analysis as well as the non-adiabatic resonant conversion taking account of the astrophysical uncertainties is warranted and the more quantitative numerical analysis of the axion conversion into photons near the neutron star will be presented in the future work. The future radio telescopes could open up a new avenue for the pulsar science with the unpreceded high sensitivity and the accompanying science such as the axion search discussed in this paper would deserve the further exploration \cite{kra2015}.

\section*{Acknowledgment}
This work was supported by Institute for Basic Science (IBS-R018-D1) and JSPS KAKENHI Grant (15K17646, 17H01110). We thank S. Youn and IBS Center for Axion and Precision Physics Research for drawing our attention to the radio telescope probes on the axion. 


\end{document}